\documentclass{article}
\usepackage[utf8]{inputenc}
\usepackage{color}
\usepackage{cmap}
\usepackage{amsmath}
\usepackage{amsfonts}
\usepackage{amssymb}
\usepackage{graphicx}
\usepackage{caption}
\usepackage{subcaption}
\usepackage{graphicx}
\usepackage[sort,numbers]{natbib}
\usepackage{xcolor}
\usepackage{totcount}
\usepackage{textcomp}
\usepackage{hyperref}
\graphicspath{ {./Images/} }
\usepackage{color,soul}

\usepackage{setspace} 
\title{A Survey about Acquisition System Design for Myoelectric Prosthesis}
\author{Ahmed Naguib, Dina Reda Eldamak}
\date{December 2022}

\begin{document}

\maketitle

\section{Introduction}

\begin{figure}[!ht]
\centering
\includegraphics [width=0.7\textwidth]{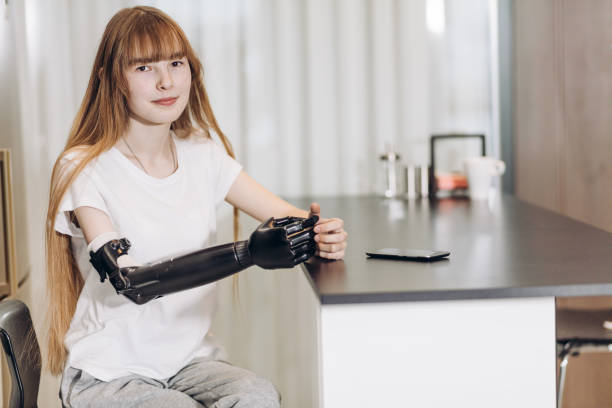}
\caption{Female with Prosthetic limb \citep{istock}}
\label{Hand_amputation}
\end{figure}

According to the World Health Organization (WHO), 30 million people are in need of prosthetic and orthotic devices  \citep{Shirley_Ryan_AbilityLab}. Some people are born with this limb loss, while others lose limbs due to diseases such as Cancer, diabetes, and work accidents. Additionally, limb amputation is among the most severe and heavily reported injuries among veterans during war \citep{UK_defence,stansbury_amputations_2008}. Example of female with hand amputation  is shown in Figure \ref{Hand_amputation}.

The medical applications of integrated circuit technology have recently made significant advances, thus improving human quality of life. Moreover, the use of microelectronics integration dominates a lot of medical applications, especially portable and wearable  battery-operated devices.
Bio-signals mostly arise from natural physiological processes, such as cardiac potentials (ECG -electro-cardiogram), potentials of the ocular tissue (EOG - electro-oculogram), potentials of the muscular tissues (electro-myogram -EMG), brain potential (electro-encephalogram -EEG), and respiratory signals , etc. 
Electro-myogram - EMG is an important factor for muscle disease diagnosis. Furthermore, it’s the key factor in connecting any amputee to a prosthetic limb. This can be done through extracting the EMG signal from the body using a readout electronics that can detect the muscles electrical activity. Consequently, the extracted signal is processed and used to control  the prosthetic limb. Thus, the objective of this report is to provide the reader with the basic understanding of integrated solutions for controlling prosthetic limbs either arms or legs.

\begin{figure}[!ht]
\centering
\includegraphics[width=5in,keepaspectratio]{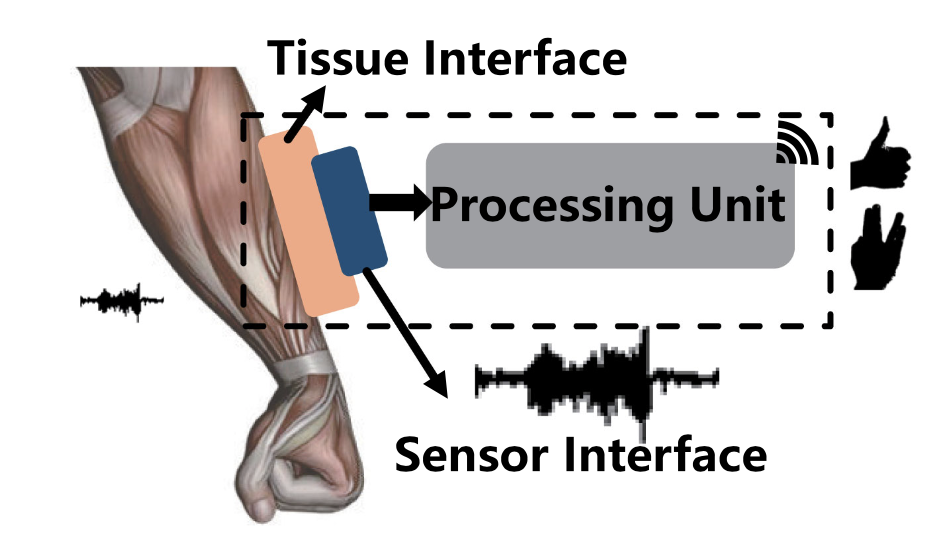}
\caption{Block diagram of a general smart sEMG recorder \citep{song2019design}}
\label{block diagram}
\end{figure}

The top level block diagram of a smart EMG acquisition system is shown in Fig.~\ref{block diagram}. The system includes a self-powered readout portable acquisition device for measuring the patient’s EMG signal in order to send it to a controller that can be used to emulate the right action to the prosthetic limb similar to the same action in a normal person.  It should be noted that miniaturized EMG acquisition system idea, which continuously monitor muscles activity, can be extended to different applications such as physical rehabilitation and prosthesis.

\section{System Architecture}

 An electronic system can control a prosethetic device by monitoring the EMG signals of the arm, and use those signals to control the prothetic arm.  Moreover, the devices can be battery-free by being powered solely using energy harvesting from the ambient.

\begin{figure}[!ht]
\centering
\includegraphics[width=\textwidth]{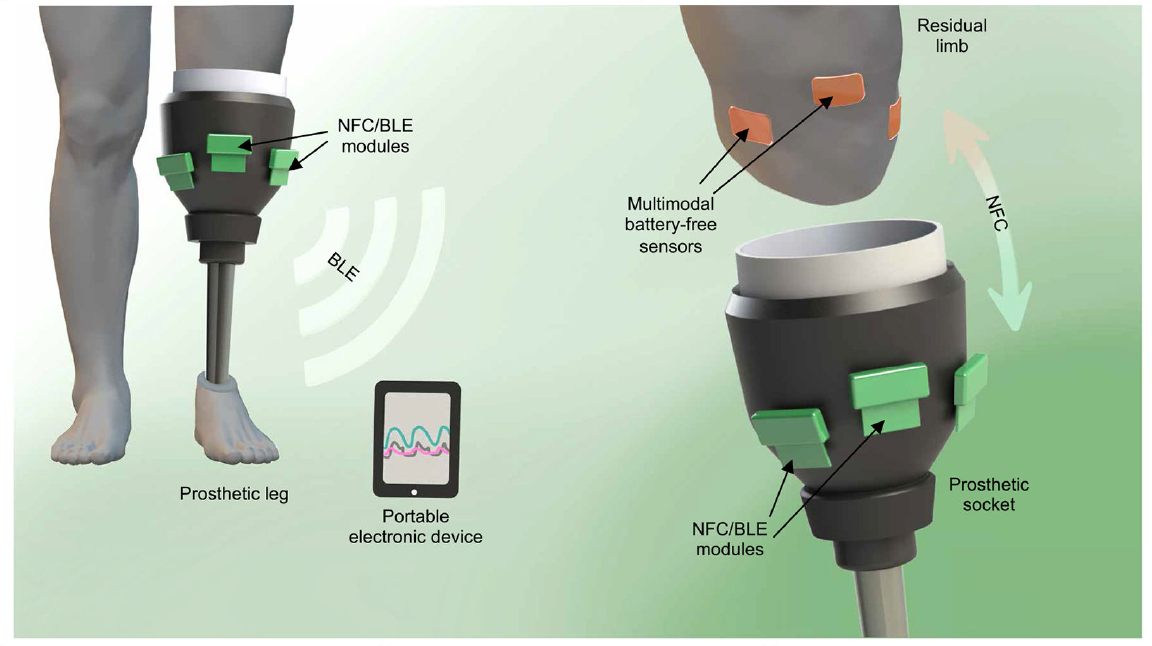}
\caption{Illustration of sensors mounted at the skin-prosthetic interface transmitting data to the device at the outer surface of the prosthetic leg using NFC and to smart phone using BLE \citep{kwak_wireless_2020-1}.}
\label{prosthetic_leg}
\end{figure}

Since these prosthetic devices requires precise fitting to the residual limb, pressure and temperature sensor at the skin-prosthetic interface are added to the system. Pressure sensors are needed for monitoring the prosthetic limb to avoid the development of regions of high pressure as the limb moves during walking or grasping objects. Temperature sensor are necessary as high temperature can accelerate tissue damage \citep{kwak_wireless_2020-1}. The signals from  the sensor at the skin-prosthetic can be transmitted to the outer surface of the prosthetic socket using Near Field Communication (NFC) or to a smart phone using Bluetooth Low Energy (BLE) as shown in Figure \ref {prosthetic_leg}.

\section{Block Diagram}

\begin{figure}[!ht]
\centering
\includegraphics[width=5in,keepaspectratio]{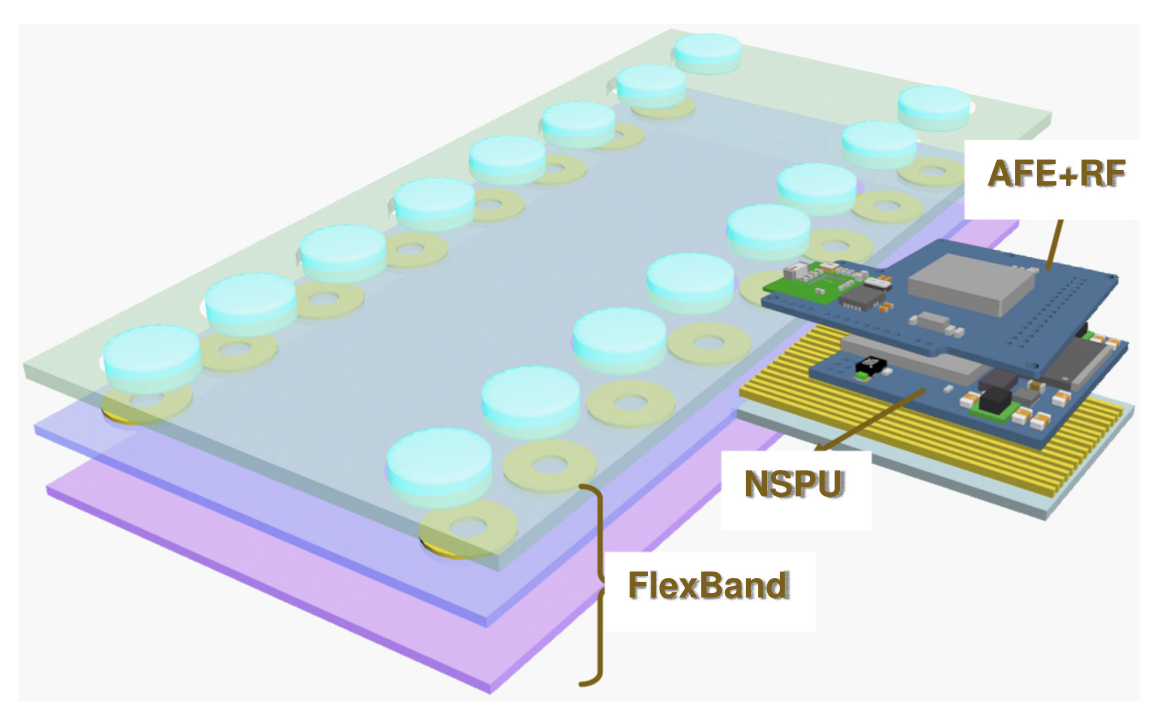}
\caption{ Block diagram of the proposed smart sEMG recorder including sensors, AFE, and RF integrated system \citep{song2019design}}
\label{sensor+AFE+RF}
\end{figure}

Three major research directions are available when designing an EMG acquisition system. The first is to acquire the signal from the surrounding noisy environment using a sensor interface circuit that’s designed in CMOS technology. The second involves reducing the form factor and power consumption of the acquisition system. The third is the signal conversion to the digital world and the interface with the digital controller. At this point, the extracted EMG signal is in a digital form and can be processed through FPGA or any other processor to control a Pprosthetic limb.

A typical block diagram of the proposed EMG acquisition system is shown in Fig. ~\ref{sensor+AFE+RF}. The system consists of an EMG sensor, analog front end (AFE), and radio frequency (RF) transmission unit. The AFE is typically composed of an analog amplification, filtration, analog to digital converter (ADC), and controller to process the digital signal and send it to a prosthetic limb. The acquisition system design can be integrated on a single chip, then the digital data is fed to FPGA or a controller.

\begin{figure}[!ht]
\centering
\includegraphics[width=\textwidth,keepaspectratio]{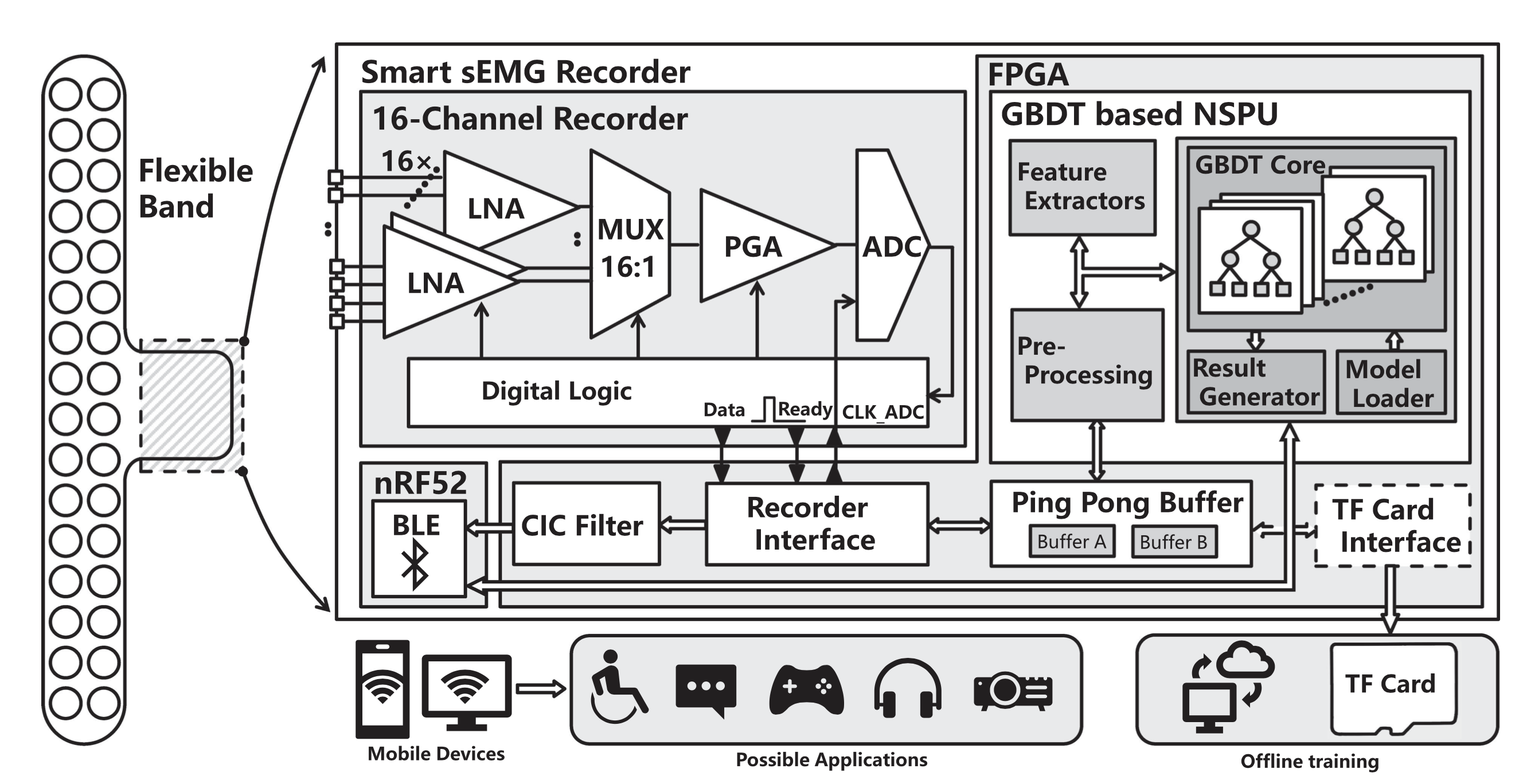}
\caption{ Detailed block diagram of the proposed smart sEMG acquisition system \citep{song2019design}}
\label{Detailed block diagram}
\end{figure}

In addition, because the integrated solution takes a considerable time during design, fabrication, and testing  phases, a discrete solution in parallel with the integrated one can be used as a proof of concept to validate the proposed methodology.

Figure ~\ref{Detailed block diagram} shows a detailed system block diagram of the proposed smart sEMG acquisition system. An analog multiplexer is inserted to choose between different EMG electrodes in the smart sEMG recorder shown in the  figure. The design of each of the building blocks involves several design challenges requiring some research. The following section includes a list of major research directions that can be pursued.

\section{Circuit Implementation}
In the following subsections, the basic system building blocks are introduced. First, the EMG sensor specifications are explored. Second, the low noise amplifier LNA design is presented. Third, the filter design and bandwidth are provided. Fourth, the signal conversion from analog to digital is presented through an ADC. Last, digital signal processing through FPGA is explored.

\subsection{Sensor Specifications}

\begin{figure}[!ht]
\centering
\includegraphics[width=5in,height=5in]{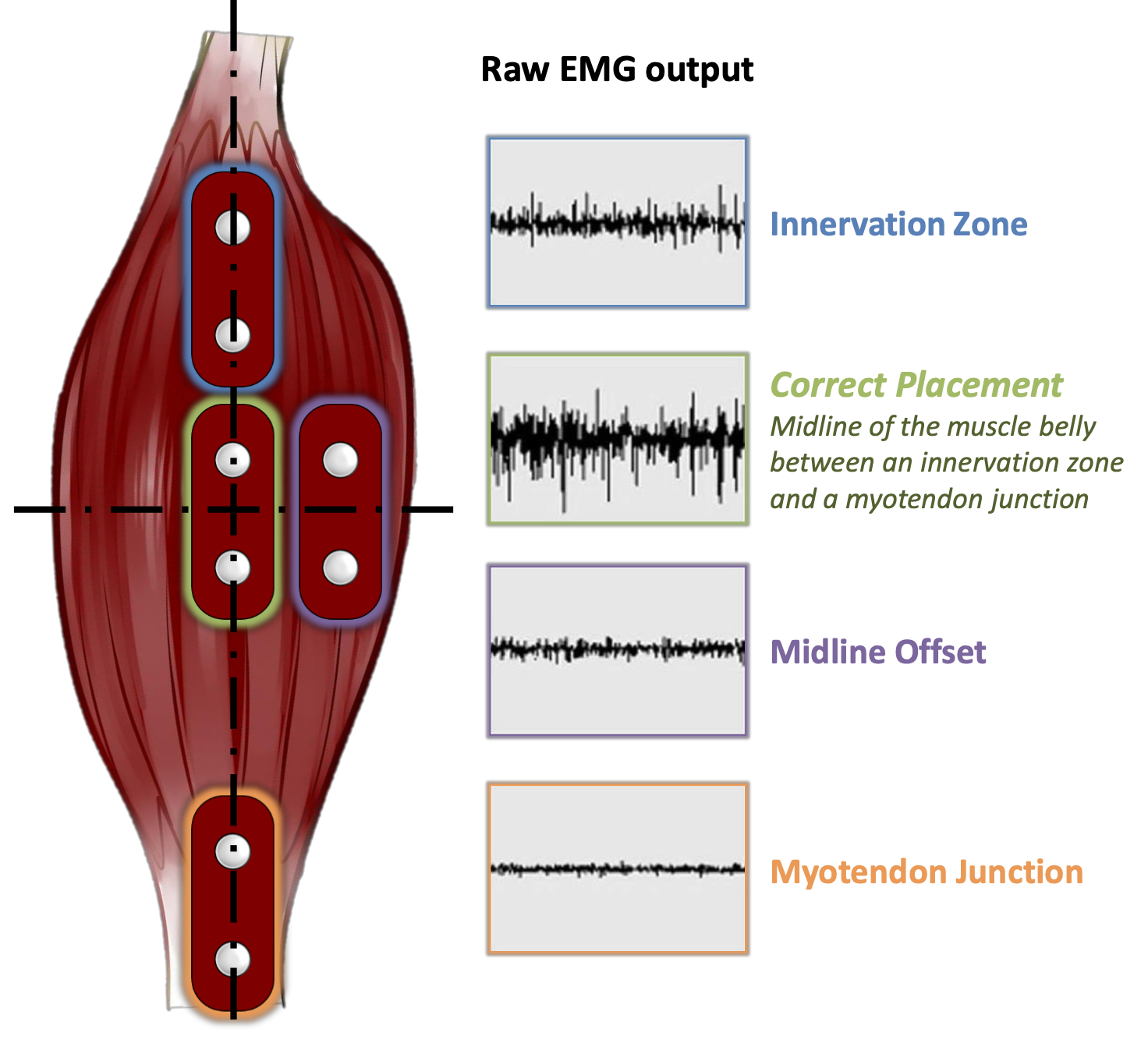}
\caption{ Effect of EMG sensor position \citep{MyoWare}}
\label{EMG sensor position}
\end{figure}

EMG sensor placement plays an important role in signal acquisition. According to its orientation and position, the EMG signal strength varies significantly. This effect is shown in Fig ~\ref{EMG sensor position}. As seen, by placing the sensor in the middle of muscle fiber, the maximum signal strength can be easily obtained. Otherwise, the signal degrades significantly when placing the sensor far away from the middle.

\begin{figure}[!ht]
\centering
\includegraphics[width=\textwidth]{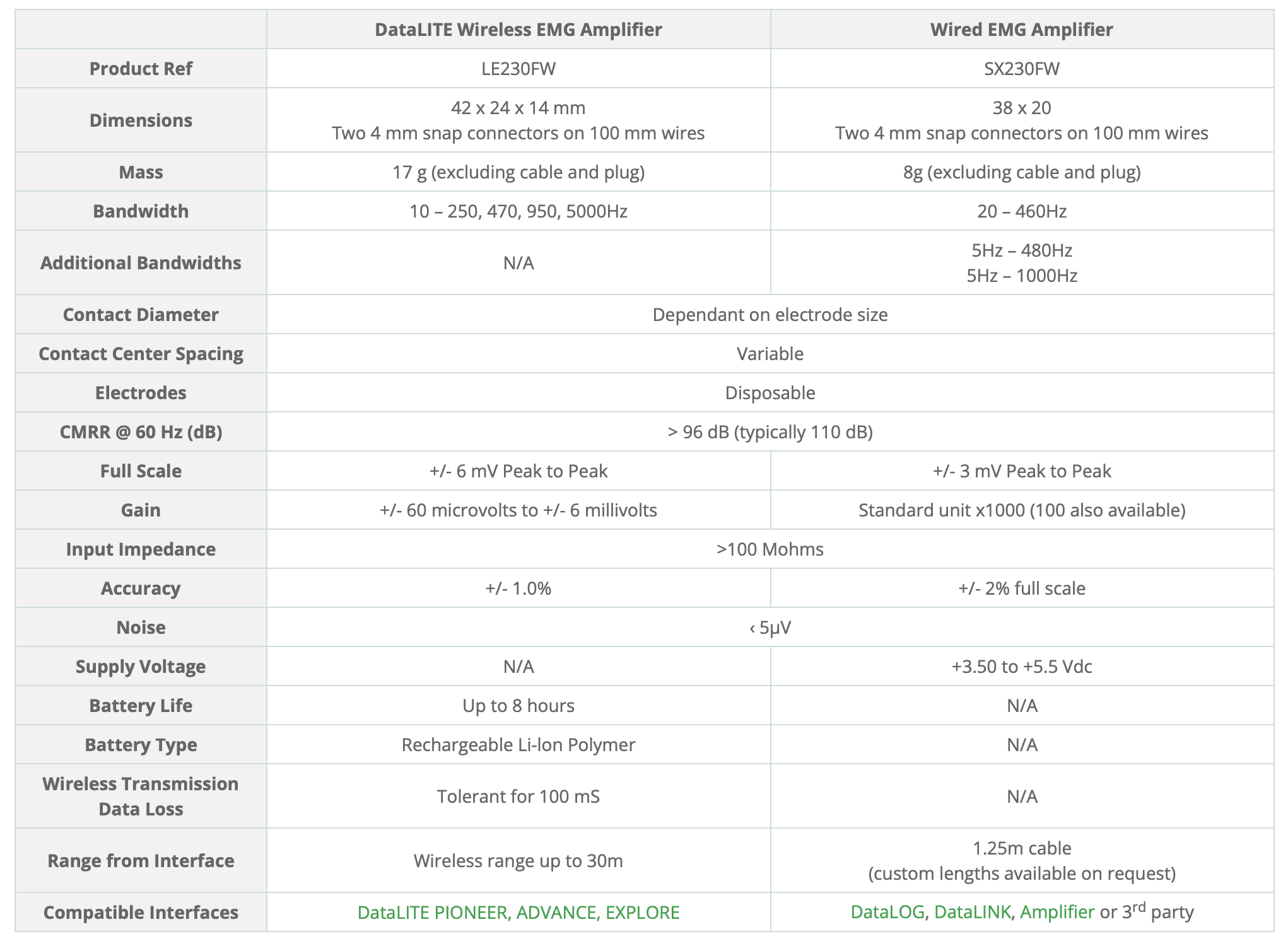}
\caption{ sensor specifications \citep{Biometric}}
\label{sensor specifications}
\end{figure}

EMG sensor can be represented in different forms. It can be in either needle that is inserted into the muscle or surface electrode that picks the signal from the skin. An example of surface EMG sensor specifications that have to be met through out the design are as follow shown in Fig.~\ref{sensor specifications}. 

\subsection{Low Noise Amplifier Design}

It's the first and the major block in the EMG chain that comes  after the sensor. The measurement sensitivity and accuracy is determined in this stage. This complicates the design and requires a large amount of adaptability to accommodate the input signal. The previous stage, which is the EMG sensor, adds large parasitic capacitance at the input of this stage, and thus reduces gain, bandwidth, noise performance and the sensitivity of the amplifier.

\begin{figure}[!ht]
\centering
\includegraphics[width=5in,height=3in]{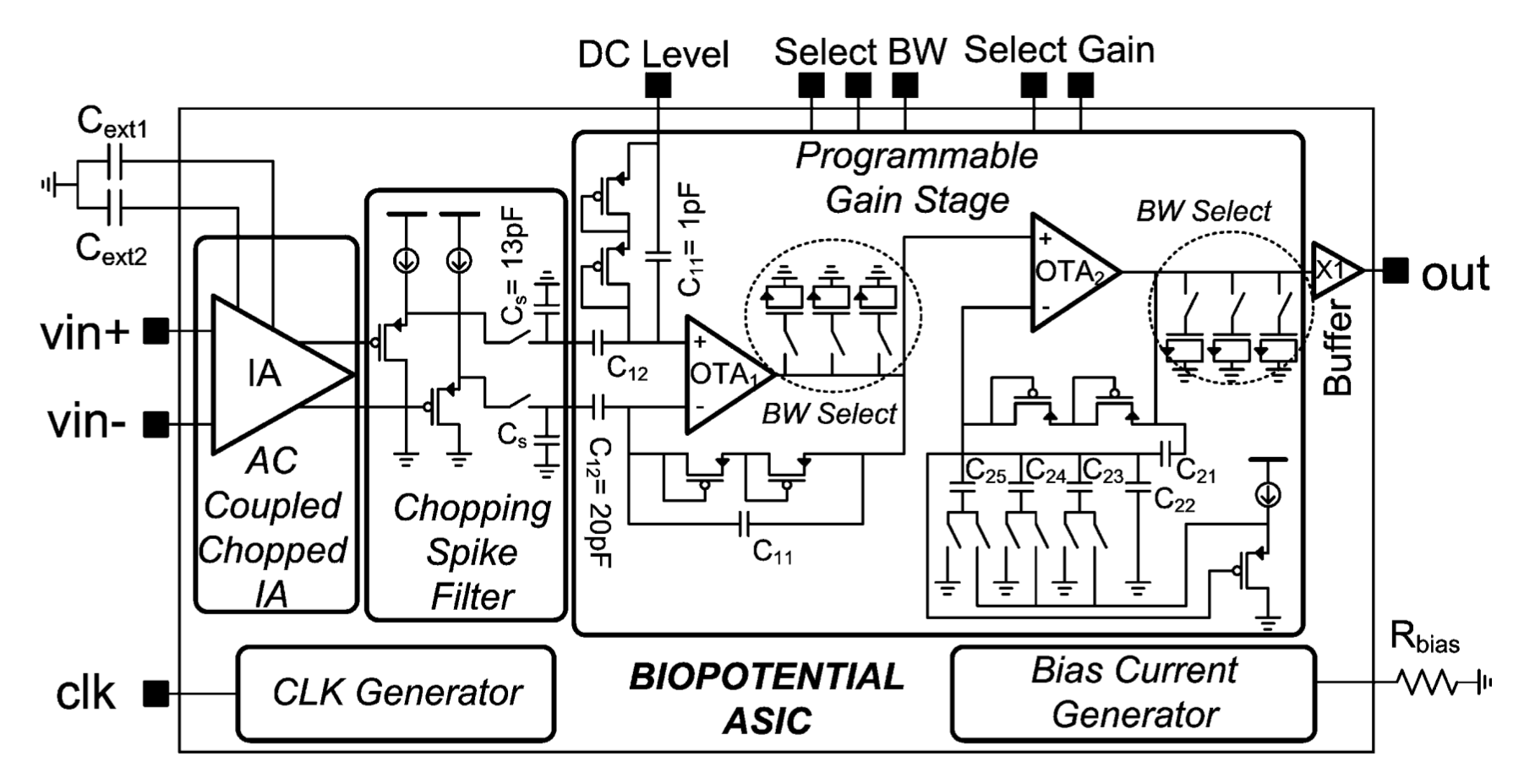}
\caption{Architecture of the bio-potential readout front-end for the acquisition of EEG, ECG, and EMG signals \citep{yazicioglu200660muw}}
\label{biopotential readout front-end}
\end{figure}

Sources of noise and interference like flicker noise, electrodes offset, and 60 Hz power line noise can affect the whole acquisition procedure. The bandwidth of the EMG signal is up to 500 Hz with amplitude that ranges from 0.1 to 5 mV and the high-frequency noise can be easily removed using a low pass filter. However, low-frequency noise and DC offset fall within the EMG bandwidth and hence require different rejection techniques. Chopping technique is one of the best candidates to modulate the offset and flicker noise to a higher spectrum which in turn enable the acquisition system to effectively suppress the interference from ambient and 1/f noise.
Different architectures with different requirements in terms of input signal levels, BW and amplitudes are proposed in literature  \cite{yazicioglu200660muw, orguc20170}. Figure ~\ref{biopotential readout front-end} shows the block diagram of implemented analog front-end for  acquiring of EEG, ECG, and EMG signals \citep{yazicioglu200660muw}. The shown diagram consists of a chopper instrumentation amplifer in addition to capacitive coupling, filter stage to remove the chopping spikes, a digitally controlled variable gain amplifier.  

\subsection{Filter Design}
\begin{figure}[!ht]
\centering
\includegraphics[width=0.5\textwidth]{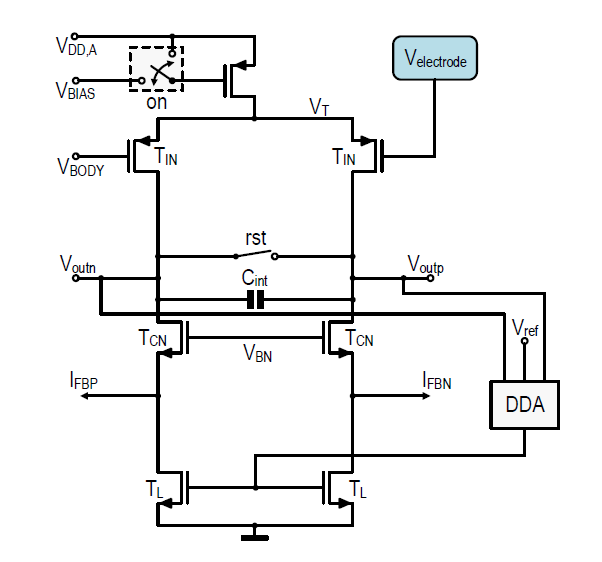}
\caption{Transistor level implementation of Gm-C filter. DDA: Differential Difference Amplifier \citep{wendler_2021}.}
\label{gm_c_filter}
\end{figure}

A Gm-C filter cab be used in the design. A standard architecture is shown in Figure \ref{gm_c_filter}. Offset from the electrodes can be canceled  using current-mode DAC \citep{wendler_2021}. Power Consumption of this topology can also be reduced by low-voltage supply operation \cite{orguc20170}.

\subsection{ADC Design}

\begin{figure}[!ht]
\centering
\includegraphics[width=0.8\textwidth]{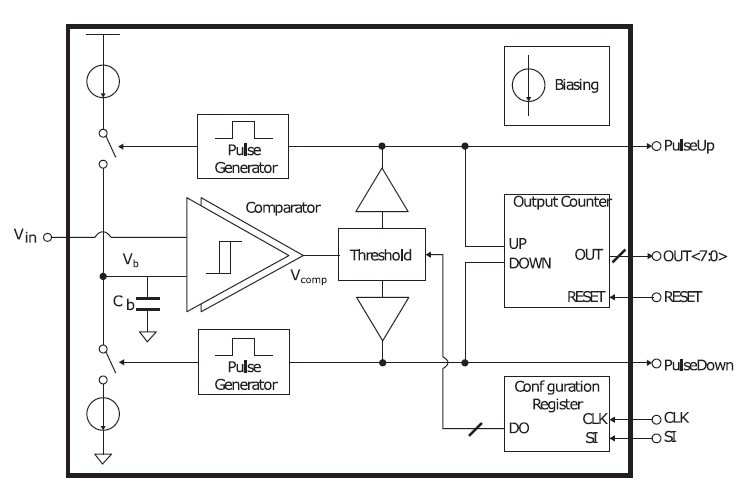}
\caption{Top level architecture of Continous-time (CT) charge based ADC with non-uniform sampling rate \citep{maslik2018}.}
\label{CT_chargebased_ADC}
\end{figure}

Non-uniform sampling can minimize the power consumption of ADC while digitizing activity-dependent biological signals. For example, a continous-time (CT) charge-based ADC that acquires samples when the input crosses a specific threshold is shown in Figure \ref{CT_chargebased_ADC}. The ADC works by storing the analog equivalent of the last digitized input as a voltage across the across the capacitor $C_{b}$. Once the input signal crosses this voltage, a pulse with length $T_P$ is generated to charge or discharge the capacitor $C_{b}$ by $V_{LSB}$ using one of the current sources connected to the supply and ground \citep{maslik2018}. Non-uniform sampling adapts to the instantaneous bandwidth of the signal, consequently the dynamic power consumption scales with the activity of the input signal. The FOM of the CT charge-based ADC can be improved by reducing the power supply further \citep{orguc20170}.

\subsection{FPGA Processing}
\begin{figure}[!ht]
\centering
\includegraphics[width=\textwidth]{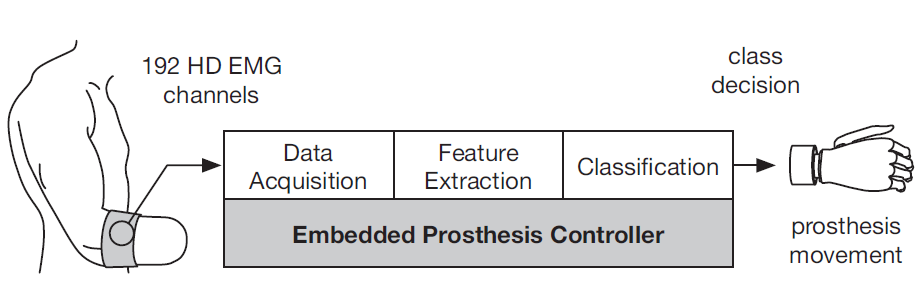}
\caption{Top level architecture of controller of prosthetic hand including feature extraction and classification  \citep{boschmann_zynq-based_2017}.}
\label{fpga_processing_emg}
\end{figure}

\begin{figure}[!ht]
\centering
\includegraphics[width=0.8\textwidth]{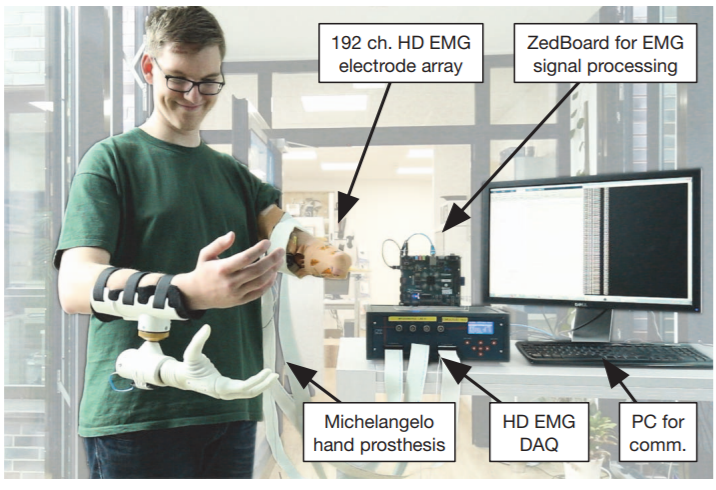}
\caption{Experimental Setup of EMG acquisition and processing using Xilinix ZedBoard \citep{boschmann_zynq-based_2017}.}
\label{fpga_setup}
\end{figure}

Machine learning algorithms such as Support Vector Machine (SVM) have allowed for on-chip feature extraction and classification of biomedical signals \citep{yoo2013}. Machine learning can also be deployed in  the domain of prosthetic devices for precise control. Figure \ref{fpga_processing_emg} depicts the controller of prosthetic device which can  be implemented using  Field Programmable Gate Array (FPGA). Figure \ref{fpga_setup} depicts the experimental setup for analyzing the data from high density EMG acquisition system  using Xilinix Zedboard \citep{boschmann_zynq-based_2017}.

\subsection{Energy Harvesting}

\begin{figure}[!ht]
\centering
\includegraphics[width=\textwidth]{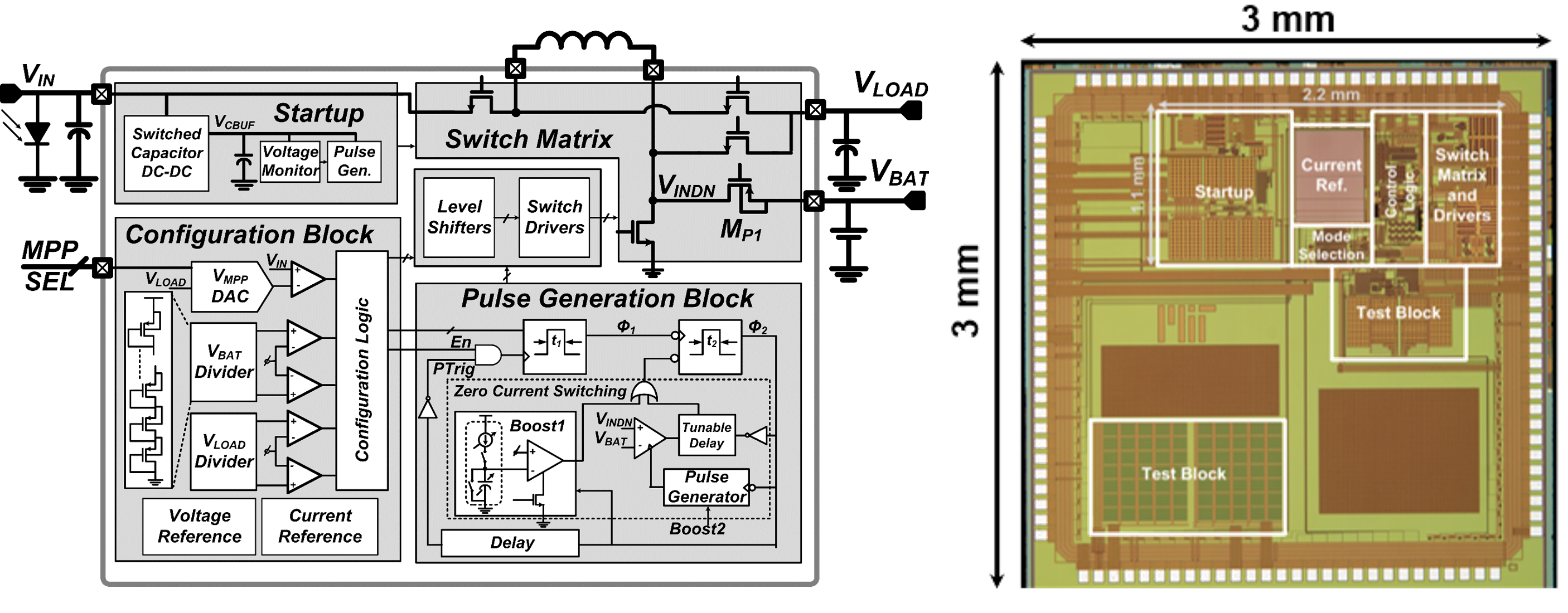}
\caption{System architecture of power management IC for solar energy harvesting applications, designed by one of the team members, and chip micrograph \citep{eldamak2016}}
\label{energy_harvesting_chip}
\end{figure}

The electrical power harvested from the environment (specially, thermal energy) can power the ultra-low-power EMG Sensor. We have previously developed energy harvesting systems from various sources and high-efficiency DC-DC converters \citep{eldamak2016, garcha2017}. For example, the system architecture of power management IC for solar energy harvesting applications , designed by the author, and chip micrograph are shown in Figure \ref{energy_harvesting_chip}. 

\section{Conclusion}
This paper provided a survey about EMG acquisition systems for prostehtics and orthotic devices.

\bibliographystyle{IEEEtran}
\bibliography{References}

\end{document}